\documentclass[runningheads]{svmult}

\usepackage{makeidx}   
\usepackage{graphicx}  
\usepackage{subeqnar}  
\usepackage{multicol}  
\usepackage{physprbb}  
\makeindex             

\begin{document}
\title*{Star-Forming Galaxies in the Sloan Digital Sky Survey -- The View from Pittsburgh}
\titlerunning{SFGs in SDSS}
\author{Regina E. Schulte-Ladbeck\inst{1}
\and Christopher J. Miller\inst{2}
\and Ulrich Hopp\inst{3}
\and Andrew Hopkins\inst{1}
\and Robert C. Nichol\inst{2}
\and Wolfgang Voges \inst{4}
\and Taotao Fang\inst{5}}
\authorrunning{Schulte-Ladbeck et al.}
%
%
\institute{University of Pittsburgh (Pitt), Pittsburgh, USA
\and Carnegie Mellon University (CMU), Pittsburgh, USA
\and Munich Observatory \& MPE, Munich, Germany
\and Max-Planck Institut f\"ur Extraterrestrische Physik (MPE), Munich, Germany
\and CMU, now at University of California, Berkeley, USA
}

\maketitle              

\begin{abstract}
We used data from the Data Release 1 (DR1) of the Sloan Digital Sky Survey (SDSS) in order to define a catalog of about 13,000 star-forming galaxies (SFG). We discuss the results of two projects. First, we matched our catalog against the ROSAT All-Sky Survey (RASS) and catalogs of pointed ROSAT observations. We identify eight X-ray emitting star-forming galaxies; four were known previously, e.g., the famous ``most metal-poor" dwarf galaxy I~Zw~18, but another four are new identifications. The data confirm the calibration of X-ray luminosity to a star-formation rate (SFR) by Ranalli et al. (2003), and are used to derive SFRs for these SDSS SFGs. We also suggest two new candidate galaxy clusters. Second, after carefully eliminating all cases of ``shredded" SFGs, we derive total absolute blue magnitudes and ionized gas metallicities (O/H ratios). We discuss the luminosity-metallicity (L-Z) relation for galaxies of different physical size (compact, small and large depending on Petrosian half-light radius, R$_e$). We report the discovery of evolution in the L-Z relation for the redshift range from 0 to 0.3, in the sense that galaxies at higher redshifts tend to have lower O/H ratios at a given luminosity. This evolution is strongest for large galaxies, and weakest for compact galaxies. 
\end{abstract}

\section{Indroduction}
The CMU-Pitt Value Added Catalog (VAC) database (http://astrophysics.phys. cmu.edu/vac/) allows users to conduct research on SDSS spectroscopic galaxy data. The current public VAC contains 133,440 SDSS DR1 galaxies, of which 124,411 are unique. This enables the pursuit of projects which require very large galaxy samples. We here describe two such projects; first, the search for rare, X-ray emitting normal SFGs, and second, an investigation of the L-Z relation for galaxies of different sizes -- compact to large -- and its evolution with redshift.

\section{Sample Definition}

The sample was drawn from all objects spectrally classified as galaxies. We required that the H$\beta$ rest equivalent width of an object be larger than 5\AA\,. The threshold of 5\AA\, was set because the SDSS pipeline does not account for underlying Balmer absorption when calculating equivalent widths, and was motivated by the work of Kobulnicky et al. (1999), who recommend a statistical correction of 3$\pm$2\AA\, for underlying H$\beta$ absorption to integrated spectra of late-type galaxies. We adopted a conservative criterion for separating SFGs and AGN. We required that the galaxies fell into the SFG regime in all three diagnostic diagrams of Kewley et al. (2001). The SDSS SFGs/AGN separation is also described in Miller et al. (2003). In addition, we checked that all galaxies classified as SFGs conform to the same relation of H$\beta$ versus [OIII]$\lambda\lambda$5007 line width. The SDSS wavelength range and the SFGs/AGN separation together, limit the redshift range of the sample from about 0.02 to 0.3. We use a concordance cosmology with $\Omega_0=0.3$, $\Omega_\Lambda=0.7$, $H_0$=70~km$\;$s$^{-1}\,$Mpc$^{-1}$. 

\section{SDSS/ROSAT SFGs}

\vspace{-1.5cm}
\begin{figure*}
\centerline{
\includegraphics[width=6.2cm]{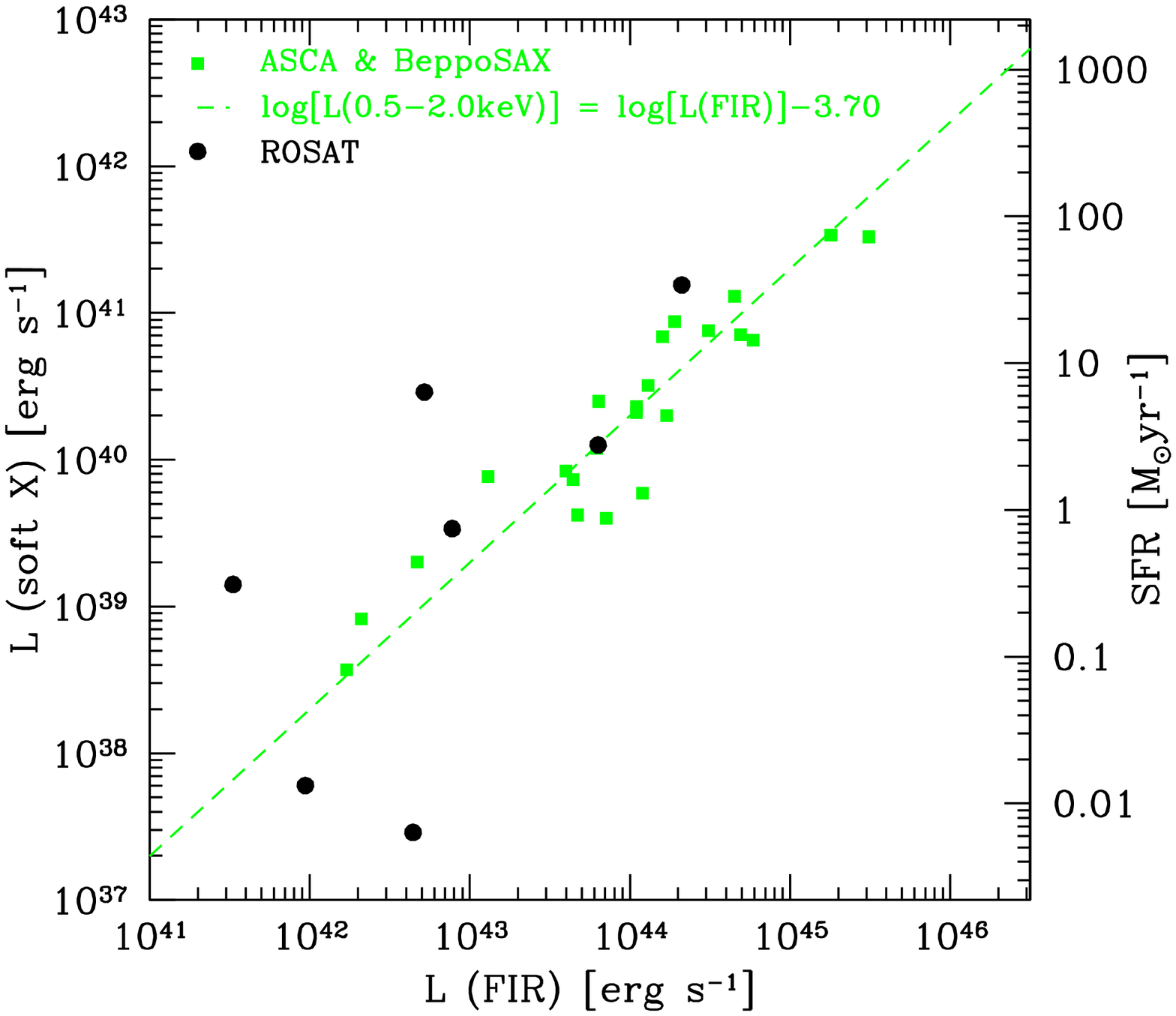}
\hspace{0.05cm}
\includegraphics[width=6.2cm]{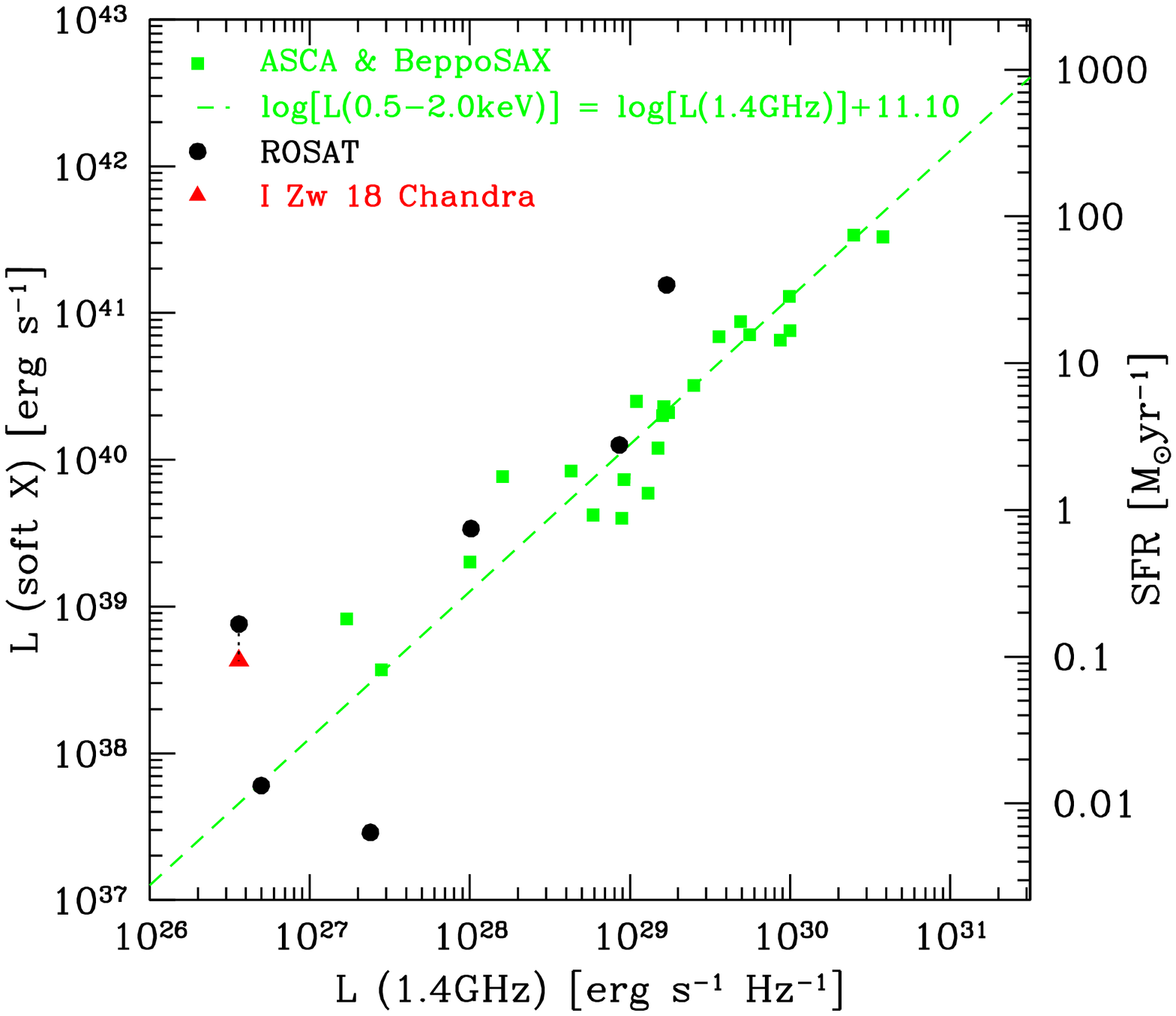}
}
\caption{{\it Left.} The soft--X-ray luminosity is shown as a function of FIR luminosity. The Ranalli et al. (2003) ASCA \& BeppoSAX data are shown by the squares; their fit, by the dashed line, and their SFR calibration, on the right-hand axis. Our ROSAT data (for the 0.5-2.0keV range) are shown by circles. {\it Right.} Same for soft--X-ray luminosity as a function of 1.4~GHz luminosity density. The triangle indicates the Chandra 0.5-2.0keV luminosity of I~Zw~18. It is lower than the ROSAT point, which had a confusing source within the extraction radius resolved by Chandra.}
\end{figure*}

We searched for positional matches to our objects with RASS Position Sensitive Proportional Counter (PSPC) data, with catalogs of pointed PSPC observations, and with catalogs of pointed High Resolution Imager (HRI) observations, maintained at MPE. The search radius was 60" for PSPC, and 20" for HRI catalogs. This resulted in a list of 146 unique SDSS DR1 SFGs which have positional coincidence with a ROSAT source (or source list). The ROSAT error circle may include a cluster of galaxies, an AGN or an X-ray bright star, as well as the SFG, but we are interested only in secure X-ray/SFG detections. AGN are X-ray bright and fairly numerous, and a match of our list against NED and SDSS QSO catalogs (Richards et al. 2002) immediately revealed that 1/4 could be QSOs, while another 1/10 could be clusters. We used a soft-band luminosity of 10$^{41}$~erg~s$^{-1}$ to separate SFGs from luminous, AGN and clusters in the remaining 2/3 of sample. In order to transform the ROSAT count rates into an unabsorbed flux, we used the Galactic HI column densities of Dickey \& Lockman (1990), and assumed an intrisic powerlaw spectrum with a photon index of 2.0. We converted redshifts to luminosity distances and used these to convert fluxes to luminosities. There are twelve objects with soft X-ray luminosities $\leq$~10$^{41}$~erg~s$^{-1}$. In addition, the data suggest two possible matches based on small f$_X$-over-f$_R$ ratios. While these have L$_X$ $>$ 10$^{41}$~erg~s$^{-1}$, they were added to our candidate sample, bringing it up to 14 objects. We investigated the properties of all 14 cases in great detail (but did not pursue the nature of the rest of the sources).

Four SFGs were previously known to be soft X-ray emitters, namely the edge-on SB galaxy IC~2233, the Blue Compact Dwarf (BCD) I~Zw~18, the SA galaxy NGC~4030, and the edge-on SA galaxy NGC~5907. There are four newly identified X-ray emitting SFGs: SDSS J014143.17+134033.16, a face-on Spiral galaxy also known as NPM1G~+13.0066, SDSS J081313.18+455940.7, the BCD Mrk~86, SDSS J130039.24+023002.81, the face-on Virgo-cluster SB galaxy NGC~4900, and SDSS J165817.7+644220.11, an inclined Spiral galaxy. The fluxes in the 0.1--2.4~keV band range from about 0.5 to 1.4$\times$10$^{-14}$~erg~s$^{-1}$~cm$^{-2}$; fairly deep because of pointed ROSAT observations. The logarithmic luminosities range from about 37.8 to 41.5. Matches coinciding with the known clusters Abell 119, Abell 318 and RDCS J0910+5422 were not attributed to SFGs. We suggest that there may be galaxy groups or clusters at J0233+0041, with a redshift of about 0.022, and at J1717+6420, with a redshift of about 0.035.

\vspace{-0.4cm}  
\begin{table} 
\caption{Optical Data}
\begin{center} 
\begin{tabular}{lcllrr}
\hline\noalign{\smallskip}
SDSS name & Other name & Type & z & d [Mpc] & $\it{r}$ [mag]$^a$\\
\noalign{\smallskip}
\hline
\noalign{\smallskip}
SDSS J014143.17+134033.16 & NPM1G+13.0066 & Spiral & 0.0453 & 201.2 & 14.84\\
SDSS J081313.18+455940.7 & Mrk 86 & BCD & 0.0015 & 6.3 & 13.31\\
SDSS J081358.8+454440.61 & IC 2233 & SB &  0.0019 & 8.2 & 13.56\\
SDSS J093402.03+551427.88 & I Zw 18 & BCD & 0.0026 & 11.0 & 16.44\\
SDSS J120026.27-010607.95 & NGC 4030 & SA & 0.0051 & 21.6 & 15.88\\
SDSS J130039.24+023002.81 & NGC 4900 & SB & 0.0032 & 13.5 & 11.81\\
SDSS J151602.21+561733.02 & NGC 5907 & SA & 0.0015 & 6.3 & 13.86\\
SDSS J165817.7+644220.11 & ... & Spiral & 0.0160 & 69.2 & 15.81\\
\hline
\end{tabular}
\end{center}
$^a$ Petrosian magnitudes; beware, I Zw 18 is ``shredded" \& magnitude under-estimated 
\label{Tab1}
\end{table} 

\vspace{-0.4cm} 
Table~1 lists the basic data for the X-ray emitting SFGs identified in this work. Table~2 gives fluxes and luminosities which can be calibrated to a SFR. Our objects follow the soft-X-ray---FIR and soft-X-ray---radio relations derived by Ranalli et al. (2003). They propose a soft-X-ray---SFR calibration based on 23 galaxies, which is nicely confirmed by our added sample of 8 SDSS/ROSAT/SFGs, as shown by Fig.~1. The last column of Table~2 gives the X-ray SFRs of our sample based on equation (14) of Ranalli et al. (2003). 

\smallskip
\noindent $\bullet$ Our results underscore the utility of soft X-ray fluxes to derive SFRs.

\begin{table} 
\caption{FIR, Radio and X-Ray Data}
\begin{center} 
\begin{tabular}{lcccccccr}
\hline\noalign{\smallskip}
SDSS & S$_{60\mu m}$ & S$_{100\mu m}$ & logL$_{FIR}$& S$_{1.4GHz}$ & 
logL$_{1.4GHz}$ & logF$_X$$^a$ & logL$_X$$^a$ & SFR$^a$\\
{\tiny name} & {\tiny [Jy]} & {\tiny [Jy]} &  {\tiny [erg s$^{-1}$]} & {\tiny [mJy]} &
{\tiny [erg s$^{-1}$Hz$^{-1}$]} & {\tiny [erg s$^{-1}$cm$^{-2}$]} & {\tiny [erg s$^{-1}$]} &{\tiny  [M$_\odot$yr$^{-1}$]} \\
\noalign{\smallskip}
\hline
\noalign{\smallskip}
S.J01..+13.. & 0.78 & 1.44 & 44.32 & 3.5 & 29.23 & -13.50 & 41.19 & 33.70 \\
S.J08..+455. & 3.46 & 6.78 & 41.97 & 10.5 & 26.70 & -13.90 & 37.78 & 0.01 \\ 
S.J08..+454.  & 0.81 & 1.19 & 41.52 & ... & ... & -12.78 & 39.15 & 0.29  \\
S.J09..+55.. & ... & ... & ... & 2.5 & 26.56 & -13.28 & 38.88 & 0.17 \\
           $^b$          & " & " & " & " & " & -13.54 & 38.63 & 0.09 \\
S.J12..-01.. & 17.24 & 45.50 & 43.80 & 153.9 & 28.93 & -12.65 & 40.10 & 2.75 \\
S.J13..+02.. & 5.52 & 14.00 & 42.89 & 46.7 & 28.01 & -12.81 & 39.53 & 0.74 \\
S.J15..+56.. & 11.52 & 44.17 & 42.65 & 50.5 & 27.38 & -14.22 & 37.46 & 0.01 \\
S.J16..+64..  & 0.14 & 0.36 & 42.72 & ... & ... & -13.30 & 40.46 & 6.32 \\
\hline
\end{tabular}
\end{center}
$^a$ The energy range is from 0.5--2.0 keV.\\
$^b$ $\it{Chandra}$ data of I Zw 18, fitted with a power law \& Raymond-Smith model.\\ 
\label{Tab2}
\end{table} 
 
\section{SDSS L-Z Relation}

Local galaxies are observed to conform to a luminosity-metallicity relation: luminous galaxies exhibit higher metallicites than do dwarfs (Melbourne \& Salzer 2002, and references therein). The L-Z relation is usually shown as total absolute B magnitude versus O/H ratio. The L-Z relation is increasingly being used to study galaxy evolution at intermediate (e.g. Lilly et al. 2003) and high (e.g. Pettini et al. 2001) redshifts, when galaxies were younger and smaller. 

In order to study the dependence of the L-Z relation on galaxy size as well as on redshift, we divided our SDSS DR1 SFG sample into three groups depending on $\it r$-band size: compact galaxies with 0$<$R$_e$[kpc]$\leq$2, small galaxies with 2$<$R$_e$[kpc]$\leq$5; this is the largest group and includes Milky-Way--sized objects, and large galaxies having 5$<$R$_e$[kpc]$\leq$100. Photometry in $\it g$ and $\it r$ was transformed to the B band following Smith et al. (2002); galactic foreground reddening was removed and K-corrections were applied (Blanton et al. 2003). In many cases, the SDSS pipeline magnitudes referred to HII regions within galaxies, rather than to entire galaxies. This phenomenon is also known as ``shredding". Of course, the compact galaxy sample was most severely affected by shredding, and we used in our analysis only ``parents" which had not been deblended into ``children". This has eliminated a number of real compact galaxies, e.g., the two HII regions of I~Zw~18 are listed as separated galaxies in the DR1. We used the N2 parameter (Denicol\'o et al. 2002) in order to derive O/H ratios. The N2 parameter has two advantages compared with the R23 index (Pagel et al. 1979): it is not affected by reddening, and its calibration to O abundance is single-valued. The SDSS fibers have a size of 3'' and the O/H ratios usually refer to the inner regions of the galaxies. The L-Z relation for about 5,000 SDSS SFGs with high signal-to-noise data is shown in Fig.~2.

The results of the KISS survey for $\approx$500 local SFGs (Melbourne \& Salzer 2002) indicate a flattening of the L-Z relation for dwarfs. Fig.~2 suggests that small SDSS SFGs have a flatter L-Z relations than large ones. We also see a large amount of scatter among compact, dwarf galaxies. We interpret this to indicate that these galaxies are affected more by feedback than luminous/large galaxies.

\vspace{-0.4cm}
\begin{figure*}
\centerline{
\includegraphics[width=6.2cm]{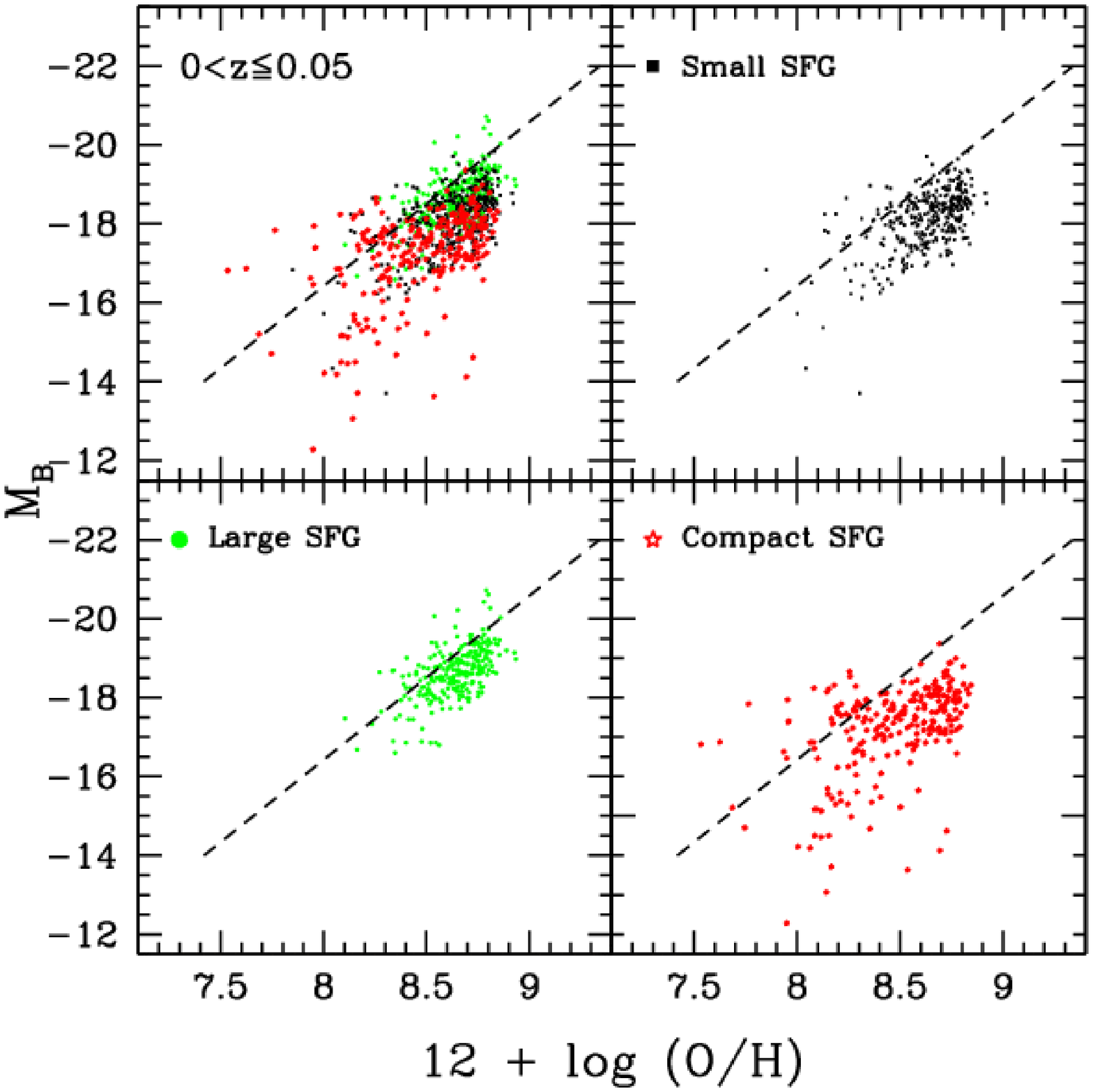}
\hspace{0.05cm}
\includegraphics[width=6.2cm]{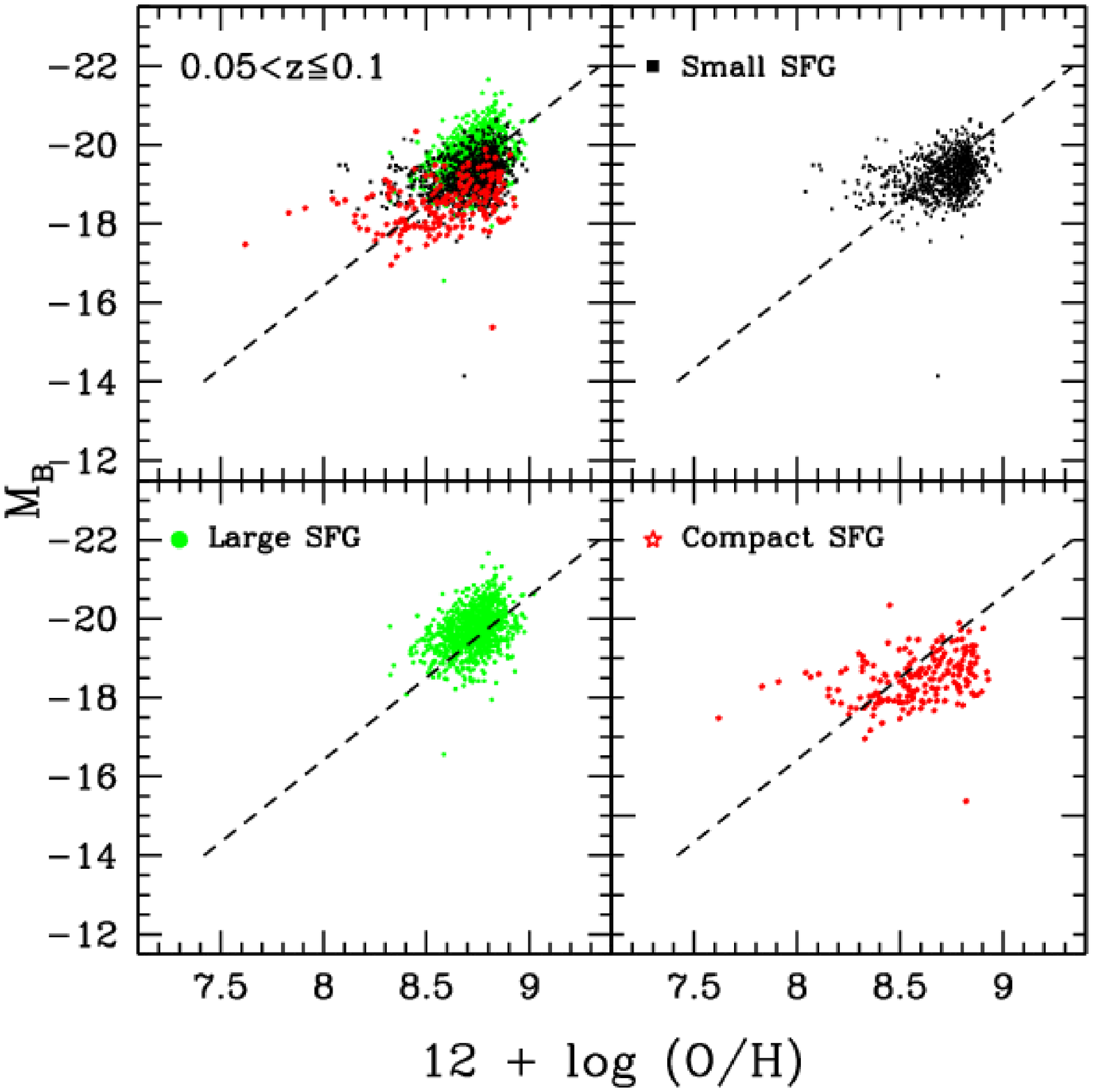}
}
\vspace{0.05cm}
\centerline{
\includegraphics[width=6.2cm]{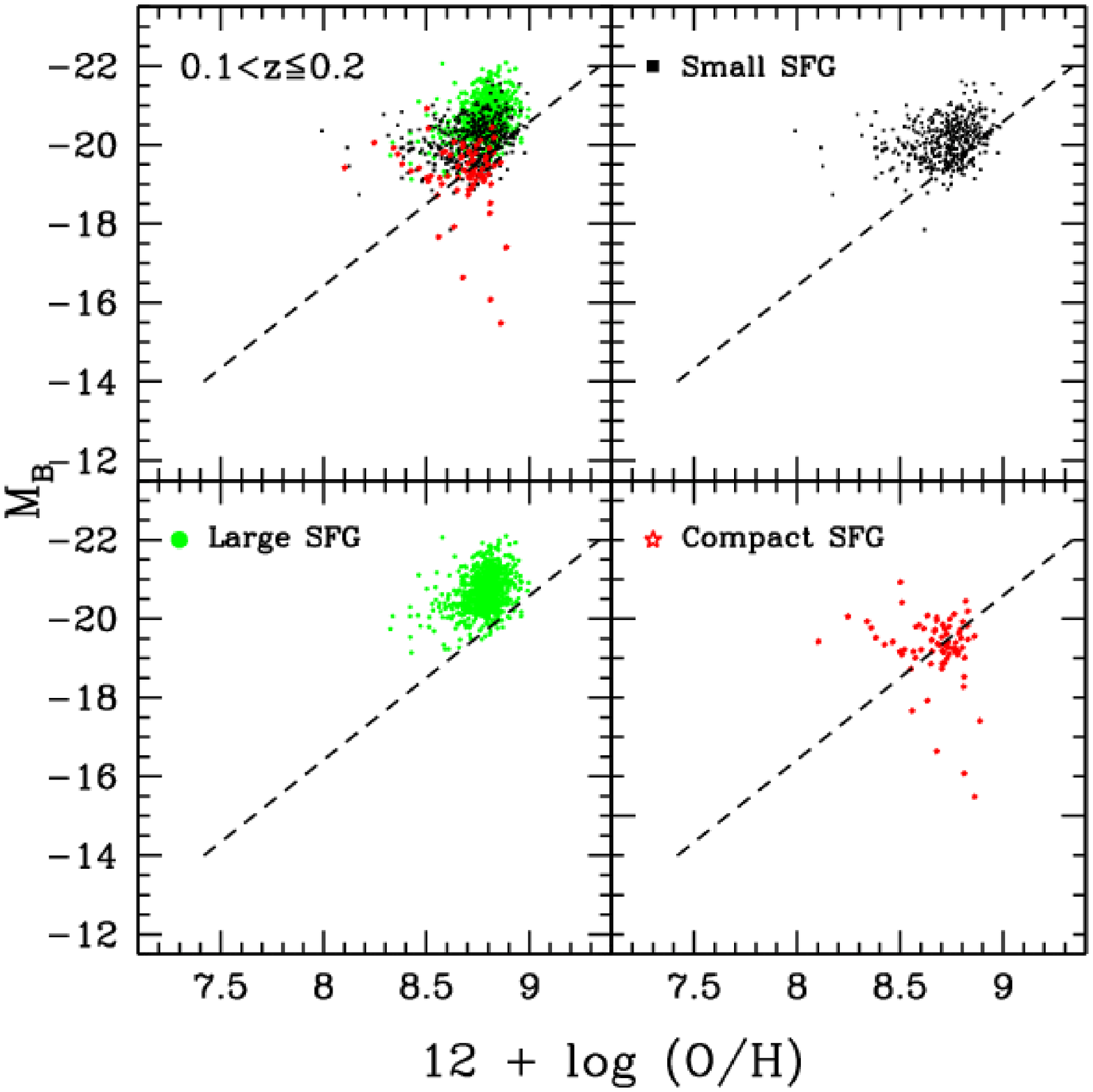}
\hspace{0.05cm}
\includegraphics[width=6.2cm]{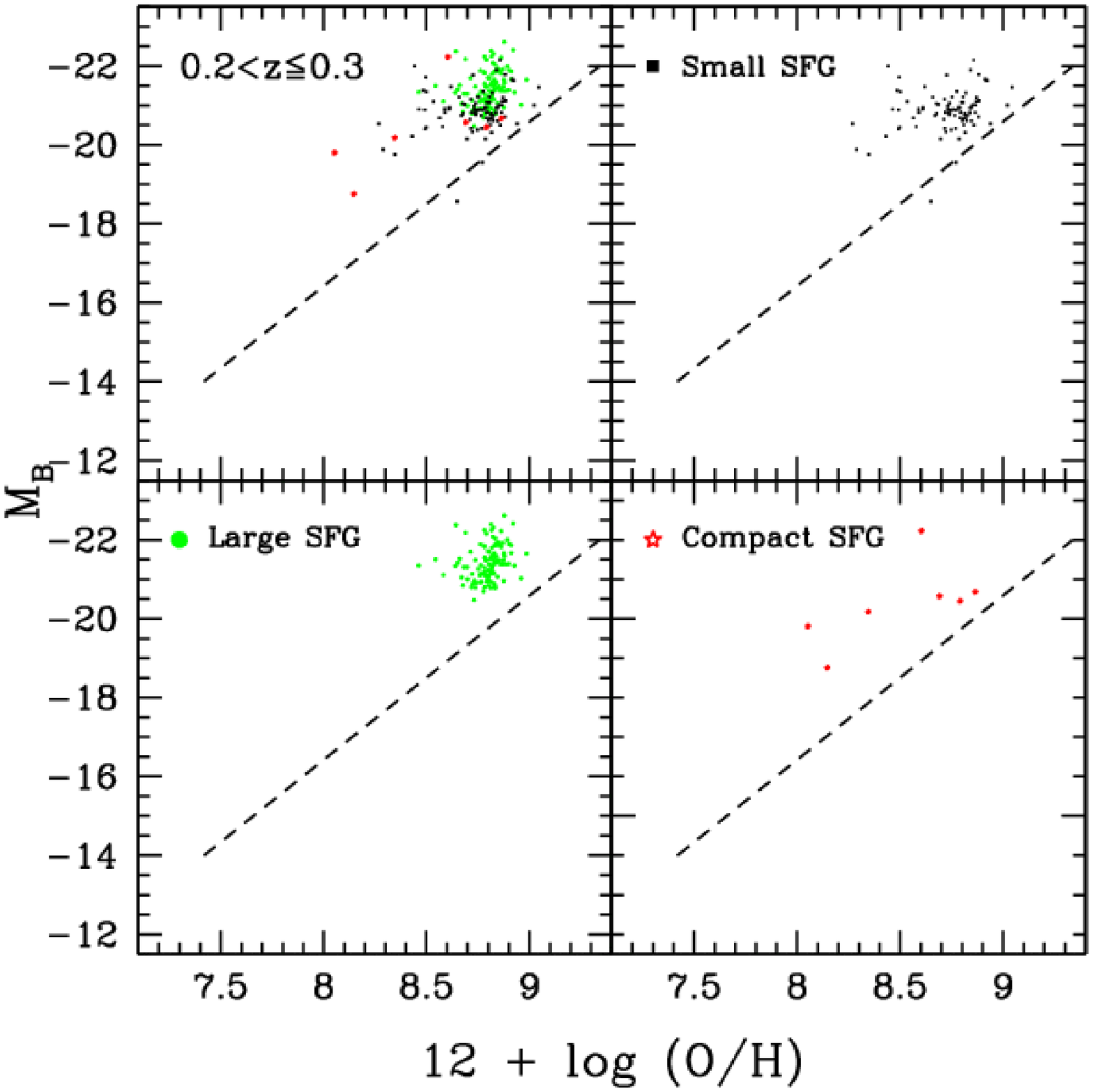}
}
\caption{The L-Z relation for SDSS SFGs in four redshift bins. The dashed line shows the relation in the KISS sample, which includes galaxies with redshifts of up to 0.095.}
\end{figure*}
\vspace{-0.3cm}

The top left quadrant of each panel in Fig.~2 shows where SDSS SFGs (of all sizes) in four redshift slices lie relative to the KISS L-Z relation (Melbourne \& Salzer 2002). We see a strong evolution, in the sense that the SDSS SFGs systematically shift across the KISS L-Z relation as redshift increases. Specifically, the data suggest that at a given metallicity (say, 12+log(O/H) = 8.7), galaxies are less luminous now than they were just 2-3 Gyrs ago; or, that at a given L (say, M$_B$ = -19.5), galaxy metallicity declines with increasing redshift.

The data in each of the redshift slices, do not extend over the luminosity range of the entire sample: in the lowest-z bin the volume is too small to include luminous galaxies, while in the highest-z bin, the limited depth of the survey precludes the detection of dwarfs. Therefore, the main concern is that luminosity introduces some kind of bias. In order to start to address this question, we used a magnitude limited sample with $r$ $\leq $17.77, as in the SDSS main galaxy sample (Strauss et al. 2002). This has the effect of removing galaxies at low O/H and decreasing the scatter among compact galaxies, but also, of cutting galaxies from the two highest-z bins. In principle, the L-Z releation could vary in both, slope and zero point, but because our galaxies cover only a small fraction of the entire L range in each redshift slice, we assumed a constant slope is retained and tested only for variation in zero point with redshift. The difference in zero point of the L-Z relation when comparing the lowest with the now highest, 0.1$<$z$\leq$0.2 bin, is significant, and amounts to over 1.5 magnitudes. 

\smallskip
\noindent $\bullet$ SDSS data indicate a very rapid evolution in the L-Z relation for late-type galaxies since z$<0$.3. 

{\bf Acknowledgments} Drs. M. Bernardi and D. Vanden Berk helped us understand SDSS parameters. RS-L thanks the MPE for the invitation to Munich in fall of 2002.  AMH gratefully acknowledges support provided by NASA through Hubble Fellowship grant HST-HF-01140.01-A awarded by the Space Telescope Science Institute (contract NAS 5-26555). Funding for the creation and distribution of the SDSS Archive has been provided by the Alfred P. Sloan Foundation, the Participating Institutions, the National Aeronautics and Space Administration, the National Science Foundation, the U.S. Department of Energy, the Japanese Monbukagakusho, and the Max Planck Society. The SDSS Web site is http://www.sdss.org/. We made extensive use of NED.


\vspace{-0.4cm}
\begin{thebibliography}{8.}
\addcontentsline{toc}{section}{References}

\bibitem{} Blanton, M.R., et al. 2003, in preparation
\bibitem{} Denicol\'o, G., Terlevich, R., Terlevich, E. 2002, MNRAS, 330, 69
\bibitem{} Dickey, J.M., Lockman, F.J. 1990, ARA\&A, 28, 215 
\bibitem{} Kewley, L. J., Dopita, M. A., Sutherland, R. S., Heisler, C. A., Trevena, J. 2001, ApJ, 556, 121
\bibitem{} Kobulnicky, H.A., Kennicutt, R.C., Jr., Pizagno, J.L. 1999, ApJ, 514, 544
\bibitem{} Lilly, S. J., Carollo, M., Stockton, A.N. 2003, astro-ph/0209243
\bibitem{} Melbourne, J., Slazer, J.J. 2002, AJ, 123, 2302
\bibitem{} Miller, C.J., Nichol, R.C., Go\',mez, P.L., Hopkins, A.M., Bernardi, M. 2003, ApJ, 597, 142
\bibitem{} Pagel, B. E. J., Edmunds, M. G., Blackwell, D. E., Chun, M. S., Smith, G. 1979, MNRAS, 198, 95
\bibitem{} Pettini, M., Shapley, A.E., Steidel, C.C., Cuby, J.G., Dickinson, M., Moorwood, A.F.M., Adelberger, K.L., Giavalisco, M. 2001, ApJ, 554, 981
\bibitem{} Ranalli, P., Comastri, A., Setti, G. 2003, A\&A, 339, 39
\bibitem{} Richards, G.R., Fan, X., Newberg, H., et al. 2002, AJ, 123, 2945
\bibitem{} Smith, J. A., et al. 2002, AJ, 123, 2121
\bibitem{} Strauss, M.A., Weinberg, D.H., Lupton, R.H. et al. 2002, AJ, 124, 1810

\end{thebibliography}
\end{document}